\documentclass[aps,prl,reprint,superscriptaddress,nofootinbib,floatfix]{revtex4-2}

\usepackage{amsmath,amssymb,amsfonts,mathrsfs}
\usepackage{graphicx}
\usepackage[colorlinks=true,linkcolor=blue,citecolor=blue,urlcolor=blue]{hyperref}

\newcommand{\ket}[1]{|#1\rangle}
\newcommand{\bra}[1]{\langle #1|}
\newcommand{\proj}[1]{|#1\rangle\langle #1|}
\newcommand{\Tr}{\operatorname{Tr}}
\newcommand{\diag}{\operatorname{diag}}
\newcommand{\id}{\mathbb{I}}

\begin{document}

\title{Entangling Atomic Quantum Memories Using High-Order Modulated Coherent States}

\author{Chaohan Cui}
\email{chaohan@umd.edu}
\affiliation{Department of Electrical and Computer Engineering, University of Maryland, College Park, MD 20742}
\author{Prajit Dhara}
\affiliation{Quantum, Photonics, and Computing Group, RTX BBN Technologies, Cambridge, MA 02138}
\author{Saikat Guha}
\affiliation{Department of Electrical and Computer Engineering, University of Maryland, College Park, MD 20742}
\affiliation{Wyant College of Optical Sciences, University of Arizona, Tucson, AZ 85721}

\begin{abstract}
High-order coherent-state modulation and collective midpoint measurements enable multiebit entanglement distribution between cavity-coupled atomic memories. An SRM-inspired 16-QAM receiver achieves $1.75$ ebits per network mode use at $0.5$~dB end-to-end loss, exceeding single-qubit-per-mode benchmarks by $5.8$~dB and lying $3.8$~dB below the half-link capacity bound. The advantage persists for per-interface loss below $0.2$~dB and calibrated phase errors below $0.2\pi$. For 4-QAM, an optimized POVM with twice as many outcomes raises the achievable rate by $4.0\%$, revealing measurement-design headroom. 
\end{abstract}

\maketitle

{\em Introduction.}---Scalable quantum information processing requires optical interconnects that distribute high-rate, high-fidelity entanglement among stationary quantum memories, processors, and sensors~\cite{Kimble2008,Wehner2018,Awschalom2021}. Distributed entanglement underpins quantum sensing and long-baseline interferometry~\cite{Komar2014,Khabiboulline2019,ZhangZhuang2021,Guo2020CVSensing}, modular fault-tolerant computing~\cite{Jiang2007,Fowler2010SurfaceCode,Monroe2014}, and quantum repeaters~\cite{Guha2015,AzumaRMP2023,Shi2025,Gu2025,Dhara2021-nl,Dhara2022-if}. A natural link-level figure of merit is therefore the number of distillable Bell pairs generated per channel use~\cite{DevetakWinter2005}, together with the number of independent spatial, polarization, and spectral modes that can be multiplexed per second~\footnote{Throughout this Letter, one \emph{network mode use} denotes one simultaneous use of a single optical mode on each Alice--Charlie and Bob--Charlie half-link. All quoted rates use this normalization; figure-axis labels of ``ebit/mode'' refer to ebits, i.e. ideal two-qubit Bell pairs, generated, per network mode use.}.

Conventional elementary links predominantly generate entanglement one qubit pair at a time. In single-photon (SP) midpoint-swap protocols, each end node creates spin--photon entanglement and the two photonic qubits are subjected to a Bell-state measurement (BSM) at the midpoint~\cite{Cabrillo1999,HongOuMandel1987,Dhara2023,Hermans2023}. Such protocols have been demonstrated across trapped ions, neutral atoms, rare-earth emitters, quantum dots, and solid-state defects~\cite{Moehring2007,Humphreys2018,Stephenson2020,vanLeent2022,Liu2024,Bernien2013,Pompili2021,Knaut2024,Ruskuc2025}, but their elementary throughput is limited because each successful attempt produces at most one Bell pair. In the single-rail (SR) and dual-rail (DR) baselines considered here, the ideal rate is at most $0.5$ ebit per network mode use; for a passive-linear-optical DR BSM, the Bell-measurement success probability further reduces the corresponding rate to $0.25$ ebit per use unless auxiliary photons or nonlinear resources are introduced~\cite{Calsamiglia2001,Ewert2014,Hilaire2023}. These probabilistic links also burden multiplexing, feed-forward, and asynchronous memory management~\cite{Collins2007Multiplexed,vanDam2017Multiplexed,Ruskuc2025,Shimizu2025GHZ}.

High-dimensional photonic encodings offer a route beyond one-pair-per-attempt operation. Photonic-qudit schemes can entangle multiple memories with a logical photonic state distributed across several modes~\cite{LoPiparo2019,Bacco2021MultiDim,Bell2022HighDim,Xie2021TimeBin,Zheng2022Qudit,Zhou2023Parallel,Xie2023OptExpress,LiuBharos2024Qudit}, but the enlarged Hilbert space is then paid for in optical modes. Continuous-variable encodings instead exploit the infinite-dimensional Hilbert space of a single bosonic mode and can approach pure-loss capacity benchmarks under idealized resources~\cite{Takeoka2014,Pirandola2017}. GKP-based protocols and optical quantum-nondemolition strategies provide powerful capacity-oriented benchmarks, but require highly nonclassical optical resources, very high processing efficiency, or strong optical nonlinearities~\cite{Winnel2022,Konno2024,DharaCapacity2025,DharaGKPMemory2025}. Coherent states, by contrast, are readily generated and displaced (in phase space) using linear optics and a coherent-state local oscillator, and can interact with spin memories through Duan-Kimble reflective cavity interfaces studied and demonstrated on several platforms~\cite{DuanKimble2004,Reiserer2014Gate,Reiserer2015,Bhaskar2020Memory,Borregaard2020OneWay,AzumaKato2012,Cui2025CSA}.

\begin{figure*}[t]
\centering
\includegraphics[width=1\textwidth]{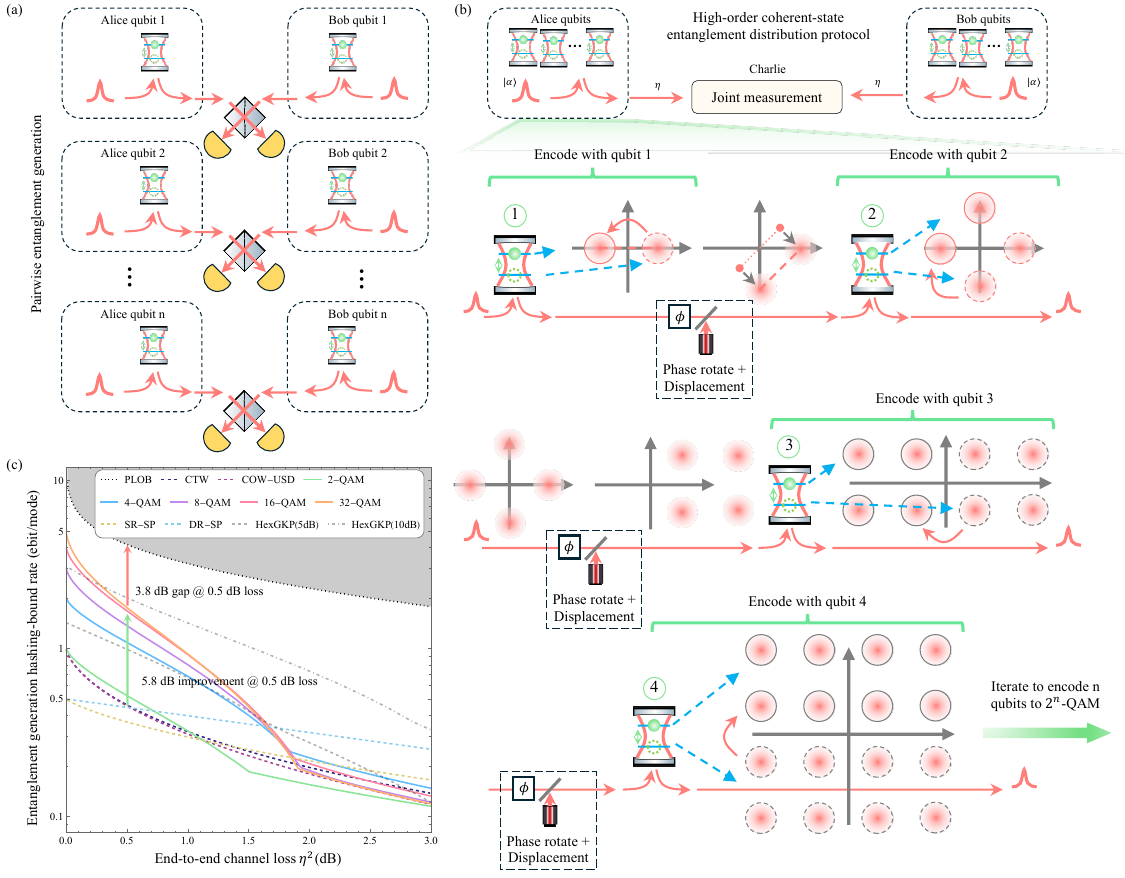}
\caption{Overview and achievable rates. (a) Conventional pairwise midpoint entanglement generation targets one remote Bell pair per elementary attempt. (b) Proposed $M$-QAM protocol: repeated cavity interactions map a $q=\log_2M$ qubit register onto an $M$-point coherent-state alphabet in one optical mode at each end node, followed by a collective two-mode measurement at Charlie. Each half-link has transmissivity $\eta$, so the end-to-end transmissivity is $T=\eta^2$ and the end-to-end loss is $L_{\rm ch}=-10\log_{10}T$. (c) Achievable hashing-bound rate per network mode use after optimizing the QAM spacing $d$. ``PLOB'' denotes $Q_2(\eta)=-\log_2(1-\eta)$, the two-way-assisted pure-loss capacity of one half-link and hence the min-cut benchmark when Charlie may implement an arbitrary quantum instrument~\cite{Pirandola2017,pirandola2019end}. SR-SP and DR-SP are the single-rail and dual-rail single-photon baselines~\cite{Dhara2023,Cui2025CSA}; CTW is the coherent two-way protocol of Ref.~\cite{Cui2025CSA} and COW-USD is the coherent one-way protocol of Ref.~\cite{Cui2025CSA}, experimentally implemented by Ref.~\cite{stas2026entanglement}. All $M$-QAM curves use the SRM-inspired achievable POVM of Eq.~\eqref{eq:yklpovm}; HexGKP curves are finite-squeezing benchmarks from Ref.~\cite{DharaCapacity2025}. Interface imperfections are treated in Fig.~\ref{fig:3}.}
\label{fig:main}
\end{figure*}

In this Letter, we use high-order coherent-state modulation to distribute multiqubit entanglement between two quantum-memory registers [Fig.~\ref{fig:main}(b)]. Alice and Bob each map a $q$-qubit register onto an $M=2^q$-point QAM alphabet in one locally prepared optical mode using repeated cavity reflections interspersed with phase-space displacements and rotations. The two optical modes propagate to a midpoint receiver, Charlie, whose collective measurement projects the remote registers into an entangled state. The objective of the receiver task is no longer coherent-state discrimination. Instead, Charlie must choose a POVM whose conditional register states maximize a probability-weighted distillable-entanglement objective, producing a nonlinear, nonconvex measurement-design problem. We construct an explicit square-root measurement(SRM)-inspired two-mode joint-detection receiver to obtain an achievable hashing rate, optimize the QAM spacing, and quantify device-level loss and phase-error tolerances. We then show for 4-QAM that a more general variationally optimized POVM exceeds the SRM-achievable rate, establishing nontrivial receiver-design headroom.

{\em Protocol.}---Alice and Bob each prepare an entangled state between a $q$-qubit register and a single optical mode. We use a strongly coupled reflective-cavity qubit--photon interface that supplies a qubit-controlled $\pi$ phase shift~\cite{DuanKimble2004,Waks2006,Sun2016,Daiss2019,Hacker2019,kato2019observation,Vaneecloo2022,Vasenin2024,Ding2024HighQ,Lau2025PhotonMemory}. Conjugating this interaction by optical phase rotations and displacements maps each computational-basis value onto a prescribed coherent-state amplitude. Repeating the operation for $q$ qubits synthesizes an $M=2^q$-point QAM alphabet with spacing $d$ in that optical mode. Writing $a,b\in\{0,\ldots,M-1\}$ for the binary register labels, the ideal source state is
\begin{equation}
\ket{\Psi_0}_{A\mathcal{A}B\mathcal{B}}
=\frac{1}{M}\sum_{a,b=0}^{M-1}
\ket{a}_A\ket{b}_B\ket{\alpha_a}_{\mathcal A}\ket{\alpha_b}_{\mathcal B}.
\label{eq:source}
\end{equation}
Here $\mathcal A$ and $\mathcal B$ denote Alice's and Bob's traveling optical modes, respectively. The association between binary labels and QAM points is a convention. We use column-major (CM) labeling throughout the main text; the repeated-reflection (RR) labeling generated directly by the sequential interaction circuit is related to it by a local permutation of each memory register, as detailed in the End Matter. CM labeling may also be generated directly using qubit-controlled displacements~\cite{kikura2026single,DharaGKPMemory2025}.

Alice and Bob send modes $\mathcal A$ and $\mathcal B$ to Charlie through identical pure-loss half-links of transmissivity $\eta$, so the end-to-end transmissivity and loss are $T=\eta^2$ and 
$L_{\rm ch}\equiv-10\log_{10}T$ in dB. By transmitting through the pure-loss channel, each coherent state transforms as
$\ket{\alpha_a}_{\mathcal A}\mapsto\ket{\sqrt{\eta}\alpha_a}_{C_{\mathcal A}}\ket{\sqrt{1-\eta}\alpha_a}_{E_{\mathcal A}}$.
Defining two-mode coherent states
$\ket{\mathscr C_{ab}}=\ket{\sqrt{\eta}\alpha_a,\sqrt{\eta}\alpha_b}_{C_{\mathcal A}C_{\mathcal B}}$
and
$\ket{\mathscr E_{ab}}=\ket{\sqrt{1-\eta}\alpha_a,\sqrt{1-\eta}\alpha_b}_{E_{\mathcal A}E_{\mathcal B}}$,
the joint state before Charlie's measurement is
\begin{equation}
\ket{\Psi}_{ABCE}=\frac{1}{M}\sum_{a,b}
\ket{a,b}_{AB}\ket{\mathscr C_{ab}}\ket{\mathscr E_{ab}}.
\label{eq:joint}
\end{equation}
Loss both reduces the distinguishability of Charlie's received coherent states and dephases the memories through label information leaked to the environment. The latter is captured by
\begin{equation}
\begin{aligned}
\Lambda_{ab,a'b'}
&\equiv\langle\mathscr E_{a'b'}|\mathscr E_{ab}\rangle\\
&=\exp\!\left[-\frac{1-\eta}{2}
\left(|\alpha_a|^2+|\alpha_{a'}|^2-2\alpha_{a'}^*\alpha_a\right)\right]\\
&\quad\times
\exp\!\left[-\frac{1-\eta}{2}
\left(|\alpha_b|^2+|\alpha_{b'}|^2-2\alpha_{b'}^*\alpha_b\right)\right].
\end{aligned}
\label{eq:env-overlap}
\end{equation}

\begin{figure}[t]
\centering
\includegraphics[width=\linewidth]{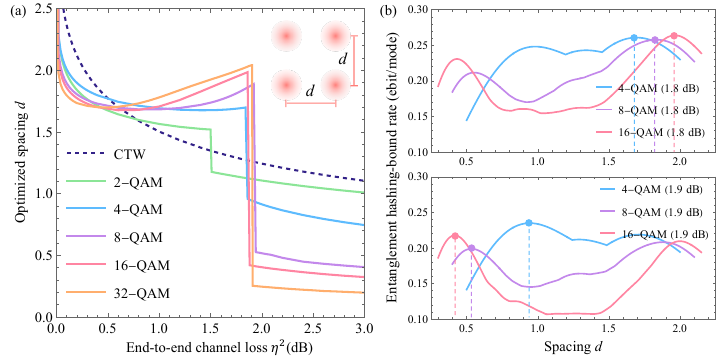}
\caption{Optimization of the QAM spacing. (a) Globally selected grid-search optimum $d$ versus end-to-end loss $L_{\rm ch}$ for several QAM orders; the CTW optimum is shown for reference. (b) Achievable hashing-bound rate versus $d$ for 4-, 8-, and 16-QAM at $L_{\rm ch}=1.8$ and $1.9$~dB. Filled markers and dashed vertical lines identify the selected maxima. The competing local maxima cause the branch switches in panel (a).}
\label{fig:2}
\end{figure}

{\em Midpoint measurement.}---Charlie's receiver is a POVM
$\{\mathcal M_\mu\}_{\mu\in\mathcal K}\cup\{\mathcal M_\emptyset\}$ on the two received optical modes, where
$\mathcal K$ contains the retained heralding outcomes and
$\mathcal M_\emptyset$ collects all discarded outcomes. Conditioned on a retained outcome $\mu$, the unnormalized register state after tracing-out the environment becomes
\begin{equation}
\widetilde\rho_{AB}^{(\mu)}=\frac{1}{M^2}\sum_{a,b,a',b'}
\Lambda_{ab,a'b'}
\langle\mathscr C_{a'b'}|\mathcal M_\mu|\mathscr C_{ab}\rangle
\ket{a,b}\bra{a',b'},
\label{eq:conditioned}
\end{equation}
with probability $p_\mu=\Tr\widetilde\rho_{AB}^{(\mu)}$ and normalized state $\rho_{AB}^{(\mu)}=\widetilde\rho_{AB}^{(\mu)}/p_\mu$. We benchmark generated entanglement with the probability-weighted two-way coherent-information (hashing) lower bound~\cite{DevetakWinter2005}, using the $B\to A$ direction throughout (without loss of generality) 
\begin{equation}
R=\sum_{\mu\in\mathcal K}p_\mu \,I(\rho_{AB}^{(\mu)})
\label{eq:rate}
\end{equation}
where $I(\rho_{AB})=S(\rho_A)-S(\rho_{AB})$, and $\rho_A =\Tr_B\rho_{AB}$.
Maximization of Eq.~\eqref{eq:rate} over \textit{all two-mode POVMs} done by Charlie is a nonlinear and non-convex optimization problem, in contrast to the optimal measurement constriction for minimum-error state discrimination~\cite{YuenKennedyLax1975}. Therefore, in the second half of this Letter, we construct an explicit achievable receiver for Charlie to herald the distribution of a decent amount of entanglement between Alice and Bob.

We first generalize the Bell basis to the $M^2$ Pauli-shifted maximally entangled register states~\cite{BennettTeleportation1993}
\begin{equation}
\ket{\Phi_{r,s}}=\frac{1}{\sqrt M}
\sum_{h\in\{0,1\}^q}(-1)^{s\cdot h}
\ket{h}_A\ket{h\oplus r}_B,
\label{eq:bellbasis}
\end{equation}
with $r,s\in\{0,1\}^q$. Relabeling Eq.~\eqref{eq:joint} by
$h=a$ and $r=a\oplus b$ gives
\begin{equation}
\begin{aligned}
\ket{\Psi}_{ABCE}
&=\frac{1}{M}\sum_{r,s}
\ket{\Phi_{r,s}}_{AB}\ket{\Xi_{r,s}}_{CE},\\
\ket{\Xi_{r,s}}_{CE}
&=\frac{1}{\sqrt M}\sum_h(-1)^{s\cdot h}
\ket{\mathscr C_{h,h\oplus r}}_C
\ket{\mathscr E_{h,h\oplus r}}_E .
\end{aligned}
\label{eq:bell-decomposition}
\end{equation}
Note that the vectors $\ket{\Xi_{r,s}}$ are not individually normalized; rather they satisfy
$\sum_{r,s}\langle\Xi_{r,s}|\Xi_{r,s}\rangle=M^2$, ensuring that
$\ket{\Psi}_{ABCE}$ remains normalized. 
Motivated by these components, we define the normalized \emph{seed} states for receiver construction:
\begin{equation}
\begin{aligned}
\ket{\chi_{r,s}}&\equiv\ket{\chi_\mu}
=\frac{1}{\sqrt{\mathcal N_{r,s}}}
\sum_{h\in\{0,1\}^q}(-1)^{s\cdot h}
\ket{\mathscr C_{h,h\oplus r}},\\
\mathcal N_{r,s}
&=\sum_{h,h'\in\{0,1\}^q}
(-1)^{s\cdot(h\oplus h')}
\left\langle
\mathscr C_{h',h'\oplus r}
\middle|
\mathscr C_{h,h\oplus r}
\right\rangle,
\end{aligned}
\label{eq:chi}
\end{equation}
where $\mu\equiv(r,s)$ labels one of the $M^2$ candidate measurement outcomes and $\mathcal N_{r,s}$ normalizes the state. The set of states $\{\ket{\chi_\mu}\}_{\mu=1}^{M^2}$ spans the Hilbert space generated by the $M^2$ candidate coherent states of Charlie's two received optical modes. In the ideal limit $\eta \to 1$ and $d \to \infty$, the states $\{\ket{\chi_\mu}\}$ become orthonormal. A projective
measurement in this basis then prepares the corresponding maximally entangled state,
$\ket{\Phi_\mu}$, and the entanglement rate approaches
$\log_2M=q$ ebits per network-mode use.

As the seed states $\ket{\chi_\mu}$ are non-orthogonal for finite $d$ or $\eta<1$, we construct an SRM-inspired POVM to benchmark the achievable entanglement rate. We define the Gram matrix for $\ket{\chi_\mu}$, $G_{\mu\nu}\equiv\langle\chi_\mu|\chi_\nu\rangle$, and $G\equiv U\diag(g_1,\ldots,g_{M^2})U^\dagger$, with the corresponding
$G^{-1/2}=U\diag(g_j^{-1/2})U^\dagger$. The zero eigenvalues of $G$ are treated by the Moore-Penrose pseudoinverse, $g_j^{-1/2} \equiv 0$ for $g_j = 0$. We then define
\begin{equation}
\ket{\varphi_\mu}=\sum_\nu(G^{-1/2})_{\nu\mu}\ket{\chi_\nu}.
\label{eq:srmvec}
\end{equation}
Starting from the SRM vectors $\ket{\varphi_\mu}$, we retain the outcomes with positive $I(\rho_{AB})$ and use the heralded POVM
\begin{equation}
\mathcal M_\mu=\proj{\varphi_\mu},\qquad
\mathcal M_\emptyset=\id-\sum_{\mu\in\mathcal K}\proj{\varphi_\mu}.
\label{eq:yklpovm}
\end{equation}
When every SRM outcome is retained, the usual square-root construction is recovered on that span. Restricting to post-selection and discarded outcomes in Eq.~\eqref{eq:yklpovm} gives the SRM-inspired heralded benchmark used below.

\begin{figure}[t]
\centering
\includegraphics[width=\linewidth]{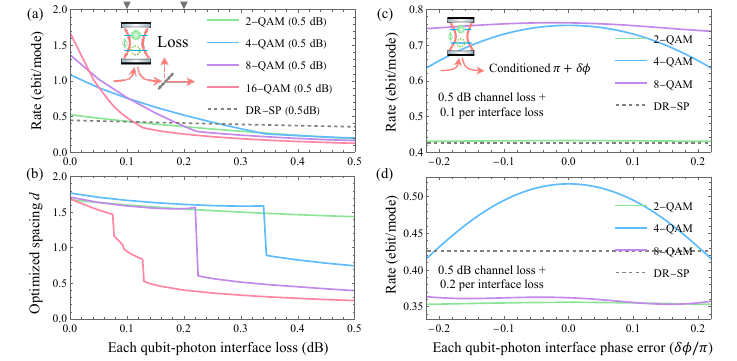}
\caption{Sensitivity to qubit--photon-interface imperfections at $L_{\rm ch}=0.5$~dB. (a) Achievable hashing-bound rate versus loss at each reflective-cavity interface. Triangles indicate $0.1$ and $0.2$~dB per-interface loss, used in panels (c) and (d), respectively. (b) Corresponding optimum QAM spacing $d$. (c),(d) Rate versus systematic conditional-phase error $\delta\phi/\pi$ at $0.1$ and $0.2$~dB per-interface loss. Charlie's receiver is constructed assuming that $\delta\phi$ has been calibrated and is known. The DR-SP baseline is shown for comparison.}
\label{fig:3}
\end{figure}

{\em Achievable rates.}---For fixed $\eta$, the SRM-inspired achievable rate depends on the nearest-neighbor QAM spacing $d$. Increasing $d$ improves Charlie's received-state distinguishability but also increases environmental label leakage and hence memory dephasing. We optimize $d$ by grid search and select the highest rate across all branches. Figure~\ref{fig:main}(c) shows the resulting rates for rectangular constellations with $M=2$ (BPSK, treated here as the two-point member of the QAM family), 4 (QPSK), 8, 16, and 32. At $L_{\rm ch}=0.5$~dB, 16-QAM on a $4\times4$ grid reaches $R=1.75$ ebits per network mode use, compared with $0.46$ ebits per use for the single-photon baselines: a factor-$3.8$ increase, or $5.8$~dB in rate ratio.

As a quantum benchmark, we use the PLOB two-way-assisted capacity of a single half-link, $Q_2(\eta)=-\log_2(1-\eta)$,
which is also the min-cut upper bound for the symmetric two-half-link pure-loss network when Charlie is allowed an arbitrary quantum instrument~\cite{Pirandola2017,pirandola2019end}. At $L_{\rm ch}=0.5$~dB, $\eta=10^{-0.025}$ and $Q_2(\eta)=4.16$ ebits per network mode use, placing the 16-QAM achievable rate $3.8$~dB below this benchmark.

Figure~\ref{fig:2}(a) shows the corresponding optimum $d$. Its discontinuities arise from competing local maxima [Fig.~\ref{fig:2}(b)]. At larger loss, a smaller-$d$ branch is favored because environmental information leakage dominates; at smaller loss, a larger $d$ is favored because Charlie benefits more strongly from increased received-state distinguishability.

The preceding curves assume ideal QAM-state synthesis. We also simulate the step-by-step implementation in Fig.~\ref{fig:main}(b), including loss and systematic phase error at every qubit--photon interaction. Interface loss can be partly precompensated by adjusting the initial amplitude and displacements to recover the target QAM grid. Nevertheless, modes lost during intermediate interactions carry history-dependent information about the register labels and cause additional dephasing. At $L_{\rm ch}=0.5$~dB, Fig.~\ref{fig:3}(a)\&(b) shows that higher-order alphabets are penalized more strongly because they undergo more sequential interactions. Figures~\ref{fig:3}(c),(d) include a systematic error $\delta\phi$ in each nominal controlled-$\pi$ phase shift. Within the calibrated-error model used here, retaining the high-order-QAM advantage requires per-interface loss $\lesssim0.2$~dB and $|\delta\phi|\lesssim0.2\pi$. A $0.2$~dB transmission penalty corresponds to approximately $95.5\%$ power throughput; in a simple overcoupled single-sided-cavity model, an external-coupling fraction of this order corresponds to $Q_i/Q_c\sim20$. These demanding targets lie in the regime pursued by recent ultra-low loss integrated photonic platforms~\cite{shen2025strong}.

\begin{figure}[t]
\centering
\includegraphics[width=\linewidth]{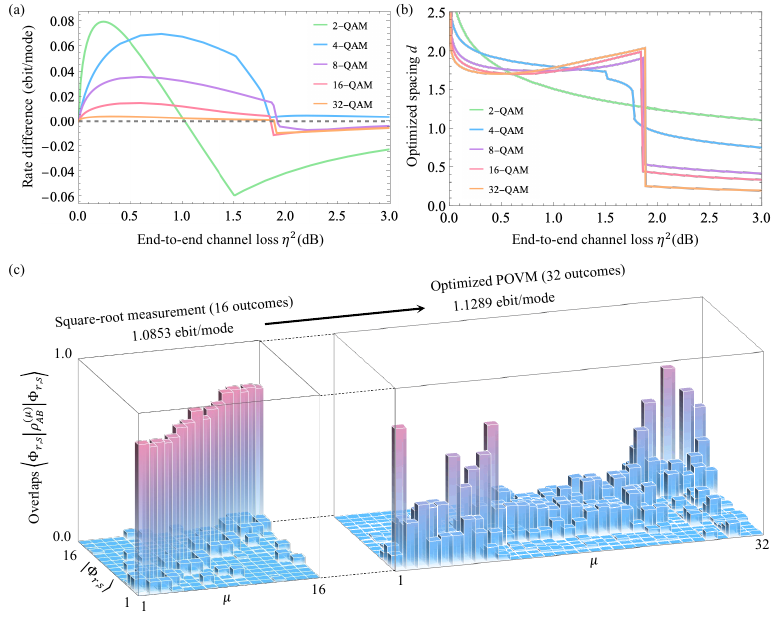}
\caption{Receiver-design headroom. (a) Rate difference $\Delta R=R-R^*$ between the full-state SRM-inspired receiver and the vacuum-omitting receiver after optimizing $d$. (b) Corresponding optimum $d$ for the vacuum-omitting receiver. (c) For 4-QAM at $L_{\rm ch}=0.5$~dB, overlaps $\langle\Phi_{r,s}|\rho_{AB}^{(\mu)}|\Phi_{r,s}\rangle$ for the 16-outcome SRM-inspired receiver and the 32-outcome variational POVM. The latter reaches $1.1289$ versus $1.0853$ ebits per network mode use; its pronounced asymmetry illustrates the nontrivial POVM-design landscape.}
\label{fig:povm}
\end{figure}

{\em Measurement-design headroom.}---The SRM-inspired receiver is a constructive achievable benchmark, not an optimizer of Eq.~\eqref{eq:rate}. Figure~\ref{fig:povm} tests two alternatives. First, we remove the joint-vacuum component from each $\ket{\chi_\mu}$, defining $\ket{\chi_\mu^*}\propto(\id-\proj{0,0})\ket{\chi_\mu}$, then construct the corresponding SRM and reoptimize $d$. The resulting rate $R^*$ is generally lower than the full-state rate $R$ in the low-loss regime [Fig.~\ref{fig:povm}(a),(b)], showing that the vacuum amplitude carries useful soft information; its small advantage at larger loss occurs outside the regime where QAM outperforms the DR-SP baseline.

Second, we variationally optimize a general 32-outcome POVM for 4-QAM at $L_{\rm ch}=0.5$~dB, jointly with $d$. We restrict this search to 4-QAM because the received-state support dimension grows as $M^2$. Multiple random initializations reduce sensitivity to local optima. The best solution raises the hashing-bound rate from $1.0853$ to $1.1289$ ebits per network mode use, a $4.0\%$ improvement. Its generalized-Bell-state overlaps are strongly asymmetric [Fig.~\ref{fig:povm}(c)], indicating that useful entanglement information spans several Bell-like sectors rather than obeying a simple symmetric discrimination rule. Because the search is nonconvex and uses a finite-outcome ansatz, this is itself only an achievable lower bound; nevertheless, it demonstrates that the SRM-inspired receiver is not hashing-optimal and that the midpoint POVM-design problem has clear performance headroom, presenting a compelling open problem for future work.

{\em Discussion.}---High-order coherent-state modulation can distribute several ebits of register entanglement using one optical mode on each half-link. In the short-reach regime studied here, rectangular QAM with an explicit SRM-inspired midpoint POVM substantially outperforms pairwise single-photon baselines while requiring only coherent states at the transmitters. Larger alphabets, however, expose the optical mode to more memory interfaces, so device loss and phase errors eventually erase the rate advantage; Figures~\ref{fig:main} and~\ref{fig:3} identify the useful operating window.

The principal open problem is the midpoint receiver: a collective, generally non-Gaussian two-mode POVM optimized for remote-register entanglement rather than coherent-state discrimination. We do not know an efficient realization using only passive linear optics, standard photonic ancillas, photon counting, and feed-forward. Existing coherent-state receivers demonstrate nonprojective measurements on overlapping alphabets~\cite{sasaki1996optimum,Becerra2015PNR,cui2022quantum}; after truncation to a finite Fock subspace, the target POVM admits a Naimark dilation~\cite{beneduci2020notes}, and hybrid continuous-variable-to-qubit transfer or continuous-variable quantum-optical storage may supply ingredients for a realization~\cite{Hastrup2022CVtoQubits,Hammerer2010Interface,Kalb2015HeraldedStorage,sajjad2026boosted}. The most compelling near-term setting is therefore a low-loss, phase-stable interconnect---for example, within a quantum data center or heterogeneous quantum-processing module---in which the multi-ebit mode-rate gain can justify the additional memory--photon interactions.

\emph{Acknowledgments.---}
This material is based upon work supported by the Defense Advanced Research Projects Agency (DARPA) under Agreement No. HR0011-26-9-E114. Approved for public release; distribution is unlimited. C.C. and S.G. also acknowledge support from the NSF Engineering Research Center for Quantum Networks (CQN, EEC-1941583). This document does not contain technology or technical data controlled under either the U.S. International Traffic in Arms Regulations or the U.S. Export Administration Regulations.

\emph{Data Availability.---} The data and codes that support the findings of this Letter are openly available at \href{https://github.com/ChaohanCui/QAM-Entanglement-Distribution}
{github.com/ChaohanCui/QAM-Entanglement-Distribution}. Codex/GPT-5.5 were used \textit{exclusively} to help organize and document the released code. All results were human-validated by the authors.

\bibliography{ref_2}

\clearpage
\section*{End Matter}

\begin{figure}
\centering
\includegraphics[width=\linewidth]{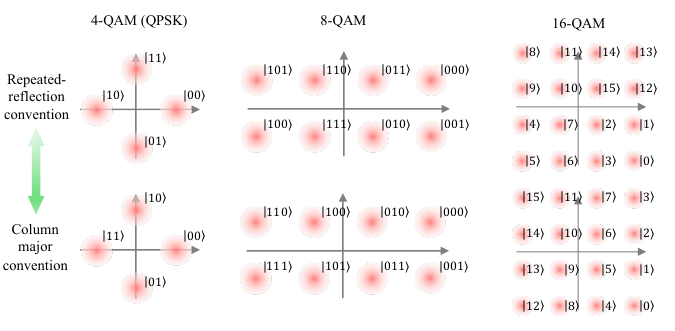}
\caption{Repeated-reflection (RR, top) and column-major (CM, bottom) labelings of the same 4-, 8-, and 16-QAM constellations. A local permutation of the memory-register basis maps one convention to the other. The 16-QAM labels are shown in decimal notation.}
\label{fig:labeling}
\end{figure}

{\em Labeling conventions.}---The sequential repeated-reflection (RR) construction naturally produces a different association between memory-register labels and rectangular-QAM points than the column-major (CM) convention used above. Figure~\ref{fig:labeling} illustrates the two orderings.

For three qubits and 8-QAM, the RR construction gives
\begin{equation}
\begin{aligned}
\ket{\Psi^{\rm RR}_{8\text{-}{\rm QAM}}}
&=\frac{1}{\sqrt8}\sum_{y_2,y_1,y_0\in\{0,1\}}
\ket{y_2y_1y_0}\ket{\alpha^{\rm RR}_{y_2y_1y_0}},\\
\alpha^{\rm RR}_{y_2y_1y_0}
&=\frac d2(-1)^{y_0}
\left[-2+(-1)^{y_2\oplus y_1}-i(-1)^{y_1}\right],
\end{aligned}
\label{eq:8qam-rr}
\end{equation}
where $d$ is the nearest-neighbor spacing in both quadratures and $\oplus$ denotes addition modulo two. The same constellation in CM order is
\begin{equation}
\begin{aligned}
\ket{\Psi^{\rm CM}_{8\text{-}{\rm QAM}}}
&=\frac{1}{\sqrt8}\sum_{x_2,x_1,x_0\in\{0,1\}}
\ket{x_2x_1x_0}\ket{\alpha_{x_2x_1x_0}},\\
\alpha_{x_2x_1x_0}
&=\frac d2\left[(2s_2+s_1)+i s_0\right],\qquad
s_j=(-1)^{x_j}.
\end{aligned}
\label{eq:8qam}
\end{equation}
One explicit local permutation $x=\pi(y)$ relating the two conventions is
\begin{equation}
x_2=1\oplus y_0,\qquad
x_1=y_2\oplus y_1\oplus y_0,\qquad
x_0=1\oplus y_1\oplus y_0.
\label{eq:labelperm}
\end{equation}
Applying the same permutation consistently to the source labeling and Charlie's measurement leaves the achievable entanglement rate invariant. Equivalently, one may retain the RR source ordering and apply the inverse permutation to the POVM labels. Changing only one convention changes the implemented measurement and can change the rate.
\end{document}